\documentclass[a4paper,11pt]{article}

\usepackage{graphicx}  
\usepackage{url}       
\usepackage{natbib}
\usepackage[margin=25mm]{geometry}
\usepackage{listings}
\usepackage{xcolor}

\title{Drawing Prolog Search Trees:\\ A Manual for Teachers and Students of Logic Programming}
\author{Johan Bos\\University of Groningen\\ \texttt{johan.bos@rug.nl}}

\definecolor{treecolour}{rgb}{0.91,0.91,0.91}
\definecolor{backcolour}{rgb}{0.95,0.95,0.95}
 
\lstdefinestyle{prolog}{
    language=Prolog,
    numbers=left,
    captionpos=b,
    frame=single,
    backgroundcolor=\color{backcolour}}

\lstdefinestyle{search}{
    basicstyle=\ttfamily\footnotesize,
    numbers=right,
    captionpos=b,
    xleftmargin=\parindent,
    backgroundcolor=\color{treecolour}}

\lstdefinestyle{rule}{
    basicstyle=\ttfamily\small,
    xleftmargin=\parindent,
    frame=single}

\begin{document}

\abovecaptionskip=-5pt 


\maketitle

\begin{abstract}
  Programming in Prolog is hard for programmers that are used to
  procedural coding. In this manual the method of drawing search trees
  is introduced with the aim to get a better understanding of how
  Prolog works. After giving a first example of a Prolog database,
  query and search tree, the art of drawing search trees is
  systematically introduced giving guidelines for queries with
  variables, conjunction, disjunction, and negation. Further examples
  are provided by giving the complete search trees that are shown in Learn
  Prolog Now!
\end{abstract}

\clearpage

\tableofcontents
\lstlistoflistings
\listoffigures

\clearpage

\section{Introduction}

Prolog, a \textit{declarative} programming language,
requires a different mindset for programmers who are used to code in
\textit{procedural} programming languages like C, Python, or Java.  It
is astonishingly hard to achieve this different attitude to coding. One approach
that sheds light on the way Prolog executes programs is to apply the brutal method of drawing
complete search trees. It is well worth the effort: mastering Prolog will
enrich the life of a programmer considerably, because it adds a
completely new (and beautiful) perspective on solving computer science
problems.

Prolog's backbone is based on a depth-first search strategy that
operates on facts and rules declared in a database
\citep{clocksinmellish:prolog,bratko,lpn}. This sounds harmless, but
it is not always straightforward to predict how a Prolog program
behaves in response to a certain query (sometimes also called
\textit{question} or \textit{goal}) given to it. The best way to find
out how Prolog reacts to a query, given a certain knowledge base, is
to draw a \textit{search tree}. In the standard logic programming
course books there is too little space devoted to drawing seach trees.

\cite{clocksinmellish:prolog}, one of the first complete introductions to
programming in Prolog, look at the way Prolog solves queries by
providing a \textit{flow of satisfaction}, which is based on the
tracing model that Prolog offers for debugging programs.
\cite{bratko} probably was the first to introduce search trees, but he
calls them \textit{execution traces}. The search trees in 
this manual are inspired by those of \citeauthor{bratko}.

Even though drawing search trees requires some effort, students of
logic programming will get a much better understanding of how their
Prolog programs work by doing so, instead of peeking at the output of
overrated tools such as the \textit{trace} facility. This manual is therefore meant
to serve as an aid in the classroom, both for students and teachers of logic
programming. Trees can be drawn in a free fashion by paper and pencil,
with the aid of sophisticated drawing software, or just by a simple
straightforward standard text characters. The latter method will be used in
this manual.

This manual is not an introduction to Prolog, and assumes the reader
has some basic experience with logic programming, and is familiar with
the concepts clause, fact, rule, predicate, backtracking, and
unification.  We will first show the basic idea of drawing search
trees in Section~\ref{sec:drawing}, where we give an example for a
simple query, a Prolog database, and the corresponding search
tree. This will give the reader a first feeling of what search trees
are and what their potential is.  In Section~\ref{sec:rules}, we show
the search tree expansion rules, for simple queries (with and without
variables), conjunctive queries, queries with disjunction, and queries
for some built-in predicates In Section~\ref{sec:negation} we show how
to draw search trees for queries with negation and the cut.  Finally,
in Section~\ref{sec:lpn} we give our version of all search trees that
occur in \cite{lpn}.






\clearpage

\section{Drawing Search Trees}\label{sec:drawing}

\subsection{The general idea}

The programming language Prolog, one of the oldest (and one of the
oddest) programming languages in the world, has its roots in
(classical) logic.  It therefore divides everything in \textit{true}
or \textit{false}.  Like classical logic, Prolog's ingredients are
based on conjunction, disjunction, implication, and negation, although
the use of these operators is restricted and Prolog's negation is
based on the closed world assumption (see Section~\ref{sec:negation}).

Prolog works by answering queries with respect to a knowledge
base. Queries are like yes/no-questions in natural language: they are
answered by Prolog as \textit{yes} (following the information in the
knowledge base), or \textit{no} (the knowledge base does not support
the query). A Prolog database comprises clauses that are either rules
or facts. Rules introduce new queries, facts lead to positive answers,
and if a piece of information cannot be deduced, it will be considered
\textit{false}. Together, the facts and rules define one or more
\textit{predicates} in the knowledge base.  Queries can also contain
variables (recall that variables are denoted by uppercase letters in
Prolog). A query with variables for which Prolog returns a positive
answer, will also instantiate the variable with values that make the
query true. Such queries are the natural language equivalent of
\textit{wh}-questions (questions starting with \textit{who},
\textit{what}, \textit{where}, and \textit{when}).

A search tree shows the way Prolog finds an answer to a query. It
shows all the steps of reasoning to come to an answer, so you may also
call it a \textit{proof tree}, using more logical terminology.  In this manual, we use the following strategy
for drawing search trees:

\begin{enumerate}

\item Write down the query in a box on the top of the page---the box
  acts as a sort of to-do list;

\item Draw the branches based on the number of clauses defined for the
  predicate that the query addresses;

\item Proceed with the branches in a left-to-right fashion. If a
  branch corresponds to a fact, go to Step 4. If a branch corresponds
  to a rule, go to Step 5.

\item If the query does not unify with the fact draw a cross
  (\texttt{x}, denoting a dead branch); otherwise (the query unifies
  with the fact) draw an empty box (\texttt{[]}, signalling that there is nothing left to do);

\item If the query does not unify with the head of the rule draw a
  cross; otherwise draw a new box with the body of the rule
  as new query. Go to Step 2.
\end{enumerate}

Of course, search trees can be drawn in many different ways. It
doesn't matter which style you're using, as long as you're consistent
(for instance, instead of the cross you can also write \textit{fail}
or \textit{no}, and instead of the empty box you can also write
\textit{yes} or \textit{success}). You can draw them with pencil and
paper, or use a handy graphical visualisation tool.

In this manual we follow, by and large, the graphical conventions
introduced in \textit{Learn Prolog Now!}  by \cite{lpn}.  It is a very
simple way to draw trees in a normal text format. However, note that
not all search trees in \textit{Learn Prolog Now!} are complete (see
Section~\ref{sec:lpn}).

Let's have look at a first example.

\subsection{A first example}

The example we consider is
taken from Learn Prolog Now! \cite[p.38]{lpn}, and describes a
situation with rather complicated love relationships between three people and a (rather naive)
definition of jealousy, as shown in Listing~\ref{listing:romance}.

\clearpage

\begin{lstlisting}[style=prolog,label=listing:romance,caption=Prolog database for a complicated romance.]
loves(vincent,mia).
loves(marsellus,mia).

jealous(A,B):- loves(A,C), loves(B,C).
\end{lstlisting}

Suppose we are interested in the answers to the query
\texttt{jealous(X,Y)}.  Following Step~1 of the recipe given above, we
start our first search tree by putting the query in an empty box (we
use square brackets to visualise boxes):

\begin{lstlisting}[style=search]
       [jealous(X,Y)]
\end{lstlisting}

This query addresses the predicate \texttt{jealous/2}. There is only
one clause defining this predicate in the database. Hence, we only
draw one branch:

\begin{lstlisting}[style=search]
      [jealous(X,Y)]
           |
           |
           |
\end{lstlisting}

Because the clause in the database is a rule, we do two things: first
we check whether the head of the rule unifies with the query. It does,
because we have for both arguments variables in the query, and
variables always unify. What happens next is that the body of the rule
in line 4 of Listing~\ref{listing:romance} will form the new query,
taking into account the variable bindings that occurred while matching
the query with the head of the rule. So this gives us:

\begin{lstlisting}[style=search]
     [jealous(X,Y)]
          |  
          | X=A
          | Y=B
          |
  [loves(A,C),loves(B,C)]
\end{lstlisting}

We now proceed at looking at this new query. It is a compound query,
and it contains two sub-queries combined with a conjunction. In this
case we first look at the left conjunct: \texttt{loves(A,C)}.  The
\texttt{loves/2} predicate is defined by two clauses in the database
as Listing~\ref{listing:romance} shows in line 1--2. So we will draw
two new branches---one for each clause:

\begin{lstlisting}[style=search]
     [jealous(X,Y)]
          |
          | X=A
          | Y=B
          |
  [loves(A,C),loves(B,C)]
       /  \ 
      /    \ 
     /      \
\end{lstlisting}

Note, at this point, that the order of the clauses in the
database---from top to bottom---corresponds to a left-to-right order
in the search tree that we draw. This is just a conventional habit
that we adopt (in a culture where one would write from right to left, we
probably would have picked a different order). But this order
\textit{is} important. As Prolog searches answers to queries by
examining clauses from top to bottom, the order of answers (in case
there is more than one answer) should be reflected in the search
tree. Hence, the order of the answers in a search tree should be read
from left to right. And this explains why we proceed with the first
branch from the left, and do not look at the second, open, branch (because
recall, Prolog's search strategy is depth-first!). The left-most branch targets
the first clause of \texttt{loves/2}, shown in line 1 of Listing~\ref{listing:romance}.
This is a fact. So the answer is
either true or false. It is true if the fact matches with the query,
which it does, because once again, the query contains only variables:
variable \texttt{A} unifies with atom \texttt{vincent}, and variable \texttt{C} unifies
with atom \texttt{mia}. 

Does this mean we have our first answer? No! We were dealing with a
compound, conjunctive, query. We did the first part of this compound
query, but still have to solve the remaining sub-query. So our new
query becomes \texttt{loves(B,mia)} and the search tree is expanded as
follows:

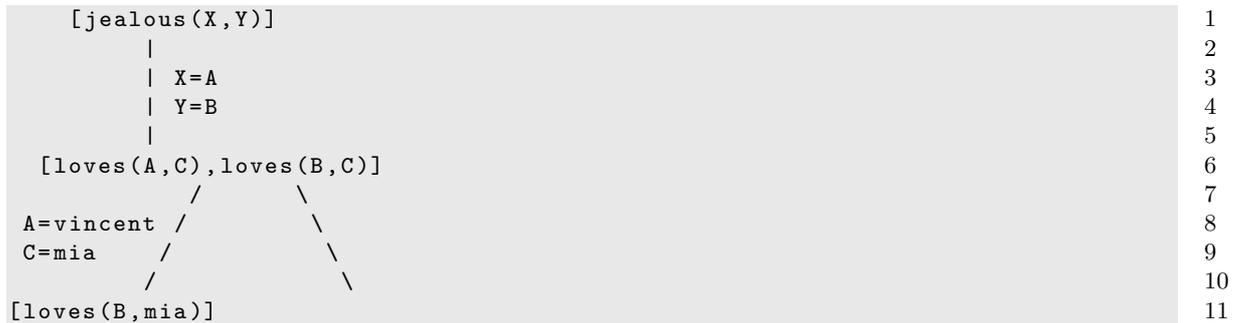
\begin{lstlisting}[style=search]
    [jealous(X,Y)]
         |
         | X=A
         | Y=B
         |
  [loves(A,C),loves(B,C)]
            /      \ 
 A=vincent /        \ 
 C=mia    /          \
         /            \
[loves(B,mia)]
\end{lstlisting}

Once more, we deal with the predicate \texttt{loves/2}, and there are
two possibilities because it is defined by two clauses, so again we
draw two new branches:

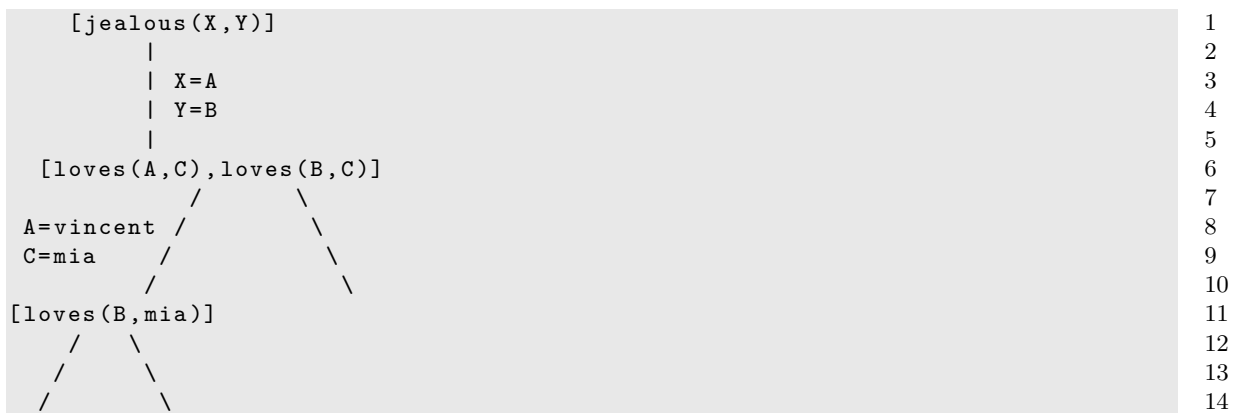
\begin{lstlisting}[style=search]
    [jealous(X,Y)]
         |
         | X=A
         | Y=B
         |
  [loves(A,C),loves(B,C)]
            /      \ 
 A=vincent /        \ 
 C=mia    /          \
         /            \
[loves(B,mia)]
    /   \
   /     \
  /       \
\end{lstlisting}

We proceed by examining the left-most branch. The corresponding fact
in the database matches with the query, and variable \texttt{B}
unifies with atom \texttt{vincent}. Because a fact does not introduce
new queries, we have our first answer, and we signify this with an
empty box in the search tree, showing that we satisfied all queries:
there is nothing more left to do on our to-do list (instead of an
empty box in line 15 you could also label the node with ``yes'' or ``true''):

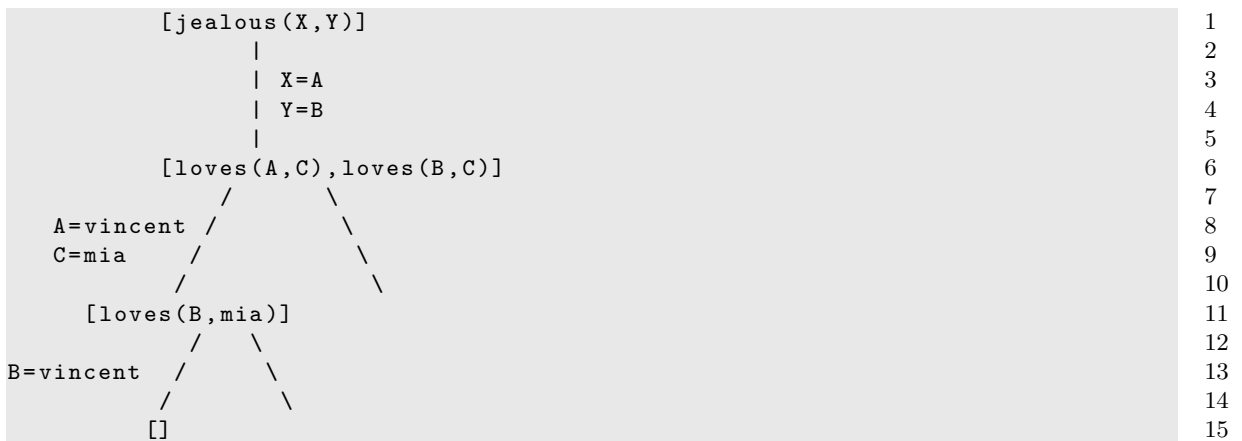
\begin{lstlisting}[style=search]
          [jealous(X,Y)]
                |
                | X=A
                | Y=B
                |
          [loves(A,C),loves(B,C)]
              /      \ 
   A=vincent /        \ 
   C=mia    /          \
           /            \
     [loves(B,mia)]
            /   \
B=vincent  /     \
          /       \
         []
\end{lstlisting}

The answer that Prolog will give to the original query is: 
\textit{true}, with
the variable \texttt{X} unified with the atom \texttt{vincent}
(because \texttt{X} unifies with \texttt{A}), and variable \texttt{Y}
unified with atom \texttt{vincent} (because \texttt{Y} unifies with
\texttt{B}). (So the answer to our original query, ``Who is jealous of whom?'', is ``Vincent is jealous of himself''. This isn't perhaps the answer you would expect, but as we have remarked before, the definition if jealousy in Listing~\ref{listing:romance} is a rather naive one and incomplete.)

\bigskip

So far so good---we got our first search tree, but it is not complete yet.
There are still open branches that might give us alternative
(positive) answers.  To make the search tree exhaustive, we perform,
like Prolog does if you ask more for solutions,
\textit{backtracking}. We go up in the tree, and find the first open
branch. We continue to expand this branch. This will give us the
following search tree and a second solution for the query:

\begin{lstlisting}[style=search]
          [jealous(X,Y)]
                |
                | X=A
                | Y=B
                |
          [loves(A,C),loves(B,C)]
              /            \ 
   A=vincent /              \ 
   C=mia    /                \
           /                  \
     [loves(B,mia)]
            /   \
B=vincent  /     \ B=marsellus
          /       \
         []       []
\end{lstlisting}

Is this search tree exhaustive? No, it isn't: there is yet another
open branch. When we turn to that, we arrive at the following search tree:

\begin{lstlisting}[style=search]
          [jealous(X,Y)]
                |
                | X=A
                | Y=B
                |
          [loves(A,C),loves(B,C)]
              /            \ 
   A=vincent /              \  A=marsellus
   C=mia    /                \ C=mia
           /                  \
     [loves(B,mia)]         [loves(B,mia)]
            /   \
B=vincent  /     \ B=marsellus
          /       \
         []       []
\end{lstlisting}

The current query that we work on the search tree, \texttt{loves(B,mia)}, is a query we have already seen before, and is
already part of the larger search tree. So we know how it will expand,
as the queries in search trees are independent from each other. This
means the complete, exhaustive search tree can be graphically realised
as follows:

\begin{lstlisting}[style=search]
               [jealous(X,Y)]
                   |
                   | X=A
                   | Y=B
                   |
              [loves(A,C),loves(B,C)]
                /                 \ 
     A=vincent /                   \  A=marsellus
     C=mia    /                     \ C=mia
             /                       \
     [loves(B,mia)]                 [loves(B,mia)]
            /   \                           /   \ 
B=vincent  /     \ B=marsellus   B=vincent /     \ B=marsellus
          /       \                       /       \
         []       []                     []       []
\end{lstlisting}

Note that variable bindings, when they appear, are decorated along the
branches of the search tree. Each Prolog answer to a query corresponds
to a path from the top query to an empty box. All the
variable bindings along this path are the instantiations used in
answering the query. For this particular query we get four different answers: 
Vincent is jealous of Vincent,  
Vincent is jealous of Marsellus,  
Marsellus is jealous of Vincent,  and
Marsellus is jealous of Marsellus.

\subsection{Next steps}

In the previous example we got acquainted with a first search tree
with a couple of exciting developments.  But there are a lot of other
things that can happen when constructing search trees.  Not all paths may
lead to an answer: they \textit{die}, instead, leading to dead
branches.  There are compound queries as we have seen, but sometimes
the queries are disjunctive instead of conjunctive, and there are
cases where we have to rename the variables.  The next section gives a
schematic overview of all cases that may appear (excluding negation
and the cut, to which we define a separate section).

\section{The Search Tree Expansion Rules}\label{sec:rules}

\subsection{A simple query without variables}

First we place the query in a box or list. Next we check how many clauses are defined
in the Prolog database for the predicate matching the query.  Then we
draw a branch for each clause that is defined by the matching
predicate. For instance, say you get a query \texttt{p(a)}, and the
predicate \texttt{p/1} happens to be defined by two clauses in the
knowledge base.  Then we can draw a  tree with two branches, and we
 concentrate on the left-most branch. There are three
possibilities: (i) there is no fact \texttt{p(a)} in the database: in this case we
draw a cross to indicate that this possibility is ``dead''; (ii) there is a fact \texttt{p(a)} in the database: we draw
an empty box to indicate we found an answer; (iii) there is a rule whose head matches the query, e.g. \texttt{p(a):- Q.}, in the
database: draw a new box for the query \texttt{Q}. All three options are
shown here:

\begin{figure}[h]
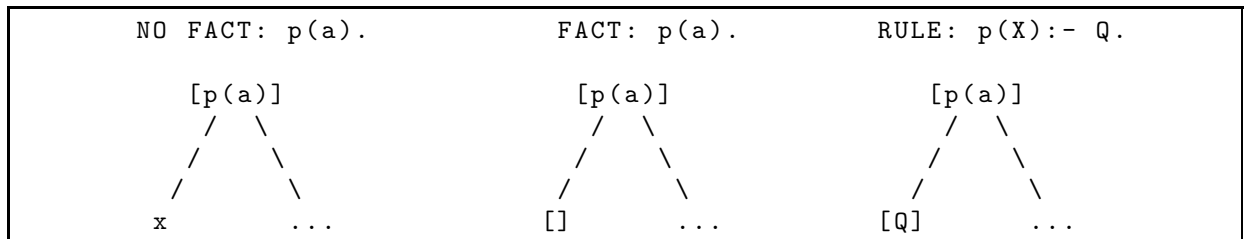

\centering
\begin{lstlisting}[style=rule]
       NO FACT: p(a).           FACT: p(a).        RULE: p(X):- Q.
     
          [p(a)]                 [p(a)]               [p(a)]
           /  \                   /  \                 /  \
          /    \                 /    \               /    \
         /      \               /      \             /      \
        x       ...            []      ...         [Q]      ...
\end{lstlisting}
\caption{Search tree expansion rules for queries without variables.}
\label{fig:ruleswv}
\end{figure}

Let's have a look at a concrete example where we can apply the rules shown in Figure~\ref{fig:ruleswv}.
Consider the following knowledge base about a small animal kingdom
with various mammals, reptiles, and birds (Listing~\ref{listing:animals}).

\begin{lstlisting}[style=prolog,label=listing:animals,caption=Prolog database for a small animal kingdom.]
mammal(fox).
mammal(whale).
mammal(bat).

bird(pelican).
bird(swan).

reptile(snake).

animal(X):- mammal(X).
animal(X):- reptile(X).
animal(X):- bird(X).
\end{lstlisting}

\clearpage

This knowledge base contains the definition for four predicates using
six facts and three rules. Suppose we want to know whether a bat is an
animal, according to this database.  The corresponding search tree for
\texttt{animal(bat)} is drawn as follows:

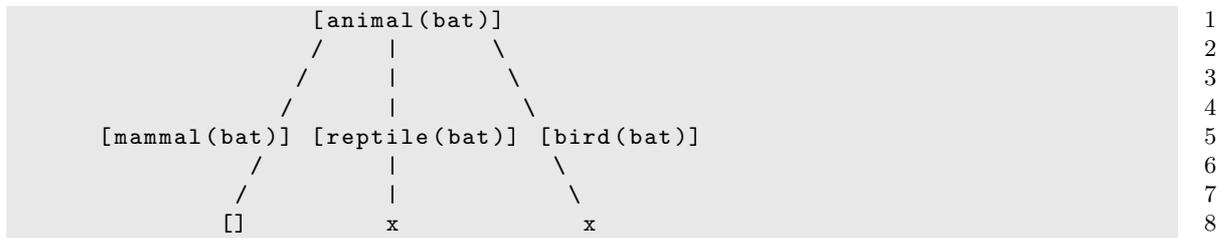
\begin{lstlisting}[style=search]
                    [animal(bat)]
                    /    |      \
                   /     |       \ 
                  /      |        \
      [mammal(bat)] [reptile(bat)] [bird(bat)]
                /        |          \
               /         |           \
              []         x            x
\end{lstlisting}

And indeed, a bat is an animal, because a bat is a mammal, and every
mammal is an animal. The search tree shows this by the branch with the
empty box in line 9.

\bigskip

What about queries that contain variables? How do the expansion rules
look for such queries? Very similar, but we need to keep track of the
variable instantiations. Let's have a look.

\subsection{Simple queries with variables}

\begin{figure}[h]
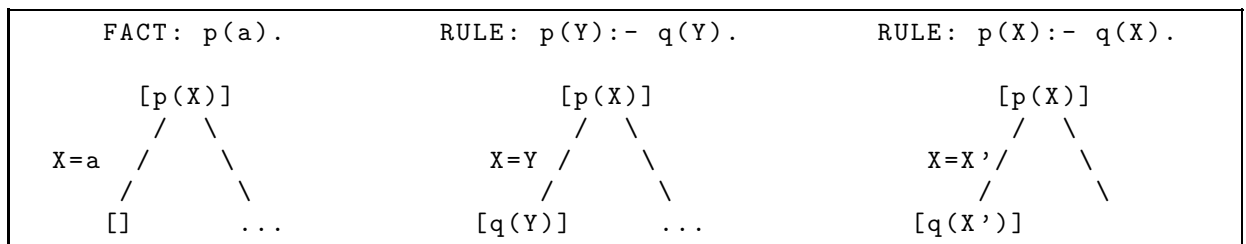

\centering
\begin{lstlisting}[style=rule]
     FACT: p(a).         RULE: p(Y):- q(Y).        RULE: p(X):- q(X).
  
       [p(X)]                   [p(X)]                    [p(X)]
        /  \                     /  \                      /  \
  X=a  /    \               X=Y /    \                X=X'/    \
      /      \                 /      \                  /      \
     []      ...           [q(Y)]     ...            [q(X')]
\end{lstlisting} 
\caption{Search tree expansion rules for queries with variables.}
\label{fig:rulesv}
\end{figure}

Here we consider search trees for queries with variables. As with any query, 
we place it in a box. Then we check how many clauses are defined
in the Prolog database for the predicate matching the query.  We
draw a branch for each clause that is defined by the matching
predicate.  But now, because there are variables in the query, each branch also needs to be annotated in case
variable bindings occur.
Like before, we need to concentrate on the left-most branch first. There are three
possibilities: 
(i) there is a fact in the database
that matches the query: in this case we draw an empty box and decorate the branch with
variable bindings; 
(ii) there is a rule whose head matches the query, for
instance \texttt{p(Y):- q(Y).}, and there no variables in the rule with the same name as in the query: add the body of the rule as a new
query to the tree;
(iii) there is a rule whose head matches the query that contains variables with the same name as in the query (e.g. \texttt{p(X):- q(X)}: in this case you need to rename the variables of the rule to avoid confusion.
All three options are shown in Figure~\ref{fig:rulesv}.

Consider again the knowledge base about the animal kingdom (Listing~\ref{listing:animals}).
The search tree for \texttt{animal(Animals)} (what is an animal?) is drawn as follows.

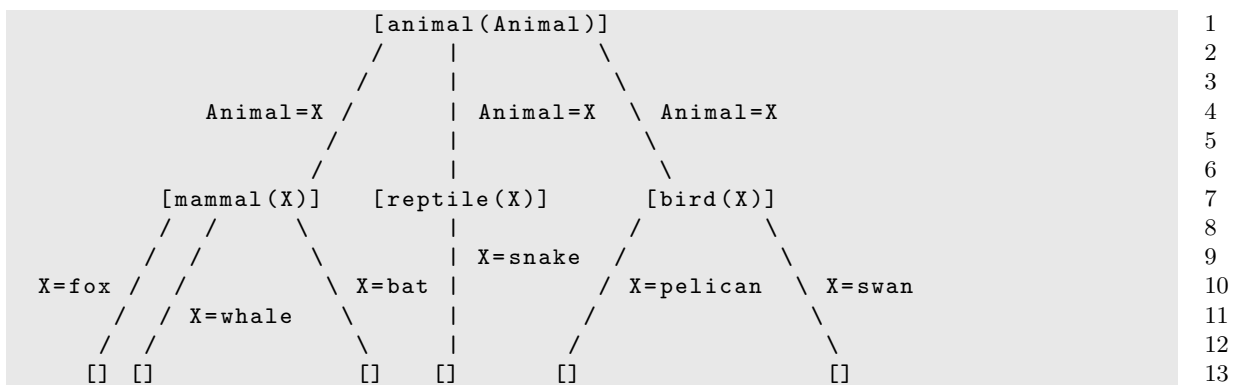
\begin{lstlisting}[style=search]
                        [animal(Animal)]
                        /    |         \
                       /     |          \
             Animal=X /      | Animal=X  \ Animal=X
                     /       |            \ 
                    /        |             \
          [mammal(X)]   [reptile(X)]      [bird(X)]
          /  /     \         |           /        \   
         /  /       \        | X=snake  /          \
  X=fox /  /         \ X=bat |         / X=pelican  \ X=swan
       /  / X=whale   \      |        /              \
      /  /             \     |       /                \
     [] []             []   []      []                []
\end{lstlisting}


\subsection{More than one query: conjunctive queries}

In case there is a conjunctive query (two or more queries separated by
a comma), we first consider the left-most query. The remaining queries
are carried over if the first one succeeds.

\begin{figure}[h]
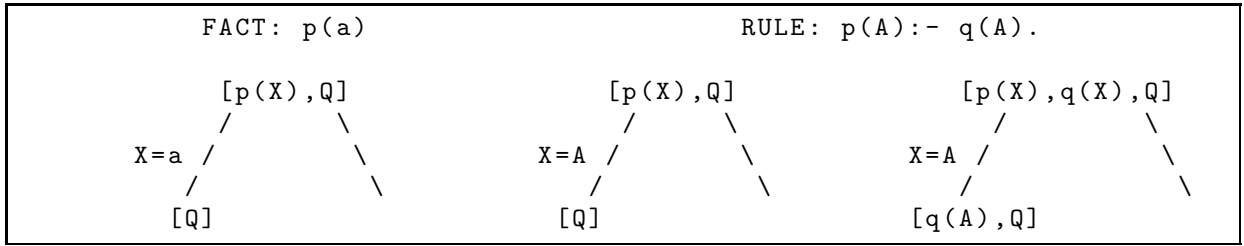

\centering
\begin{lstlisting}[style=rule]
           FACT: p(a)                      RULE: p(A):- q(A).

            [p(X),Q]               [p(X),Q]             [p(X),q(X),Q]
            /      \                /     \               /        \
       X=a /        \          X=A /       \         X=A /          \
          /          \            /         \           /            \    
         [Q]                    [Q]                  [q(A),Q]
\end{lstlisting} 
\caption{Search tree expansion rules for conjunctive queries.}
\label{fig:rulesconj}
\end{figure}

\subsection{A disjunctive query}

\begin{figure}[h]
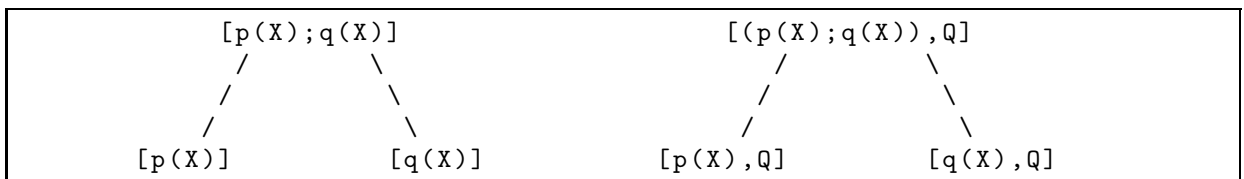

\centering
\begin{lstlisting}[style=rule]
            [p(X);q(X)]                   [(p(X);q(X)),Q]
             /       \                       /        \
            /         \                     /          \
           /           \                   /            \
       [p(X)]         [q(X)]          [p(X),Q]        [q(X),Q]
\end{lstlisting} 
\caption{Search tree expansion rules for disjunctive queries.}
\label{fig:rulesdisj}
\end{figure}

The semi-colon denotes a disjunctive query. Here it is important to
look at the scope of the disjuncts (it is good practive to use brackets to disambiguate scope ambiguities). 
Disjunction always creates two new branches, one for each
disjunct. Nothing happens to the variables, and further queries in the box are duplicated (Figure~\ref{fig:rulesdisj}).

\subsection{Lists and variable renaming}

Sometimes variables need to be renamed when drawing a search tree. 
This is required when there is a variable occuring in a query that is also used in the rule. 
We illustrate this procedure of renaming with the help of the 
 standard definition for \texttt{member/2}:

\begin{lstlisting}[style=prolog,caption=Definition of member/2]
member(X,[X|_]).
member(X,[_|L]):- member(X,L).
\end{lstlisting}

Say we follow the example in Learn Prolog Now! and have a query
\texttt{member(X,a,b,c])} \cite[p.244]{lpn}.  
The query happens to contain the variable \texttt{X}. This variable is
also used by the rule in the definition of \texttt{member/2}. To avoid
confusion we need to rename the variables in the rule each time we use
it. It is safe to do this: the scope of Prolog variables is restricted
to the rule they occur in.  So the rule above can be replaced by
\verb+member(Y,[_|L]):- member(Y,L).+, for instance.  You can pick any
(unused so far in the search tree) variable name that you like. In the
tree below we have chosen \texttt{X'}, \texttt{X''}, and so on, but we
could have also chosen \texttt{X1} and \texttt{X2} or other variables.

\clearpage

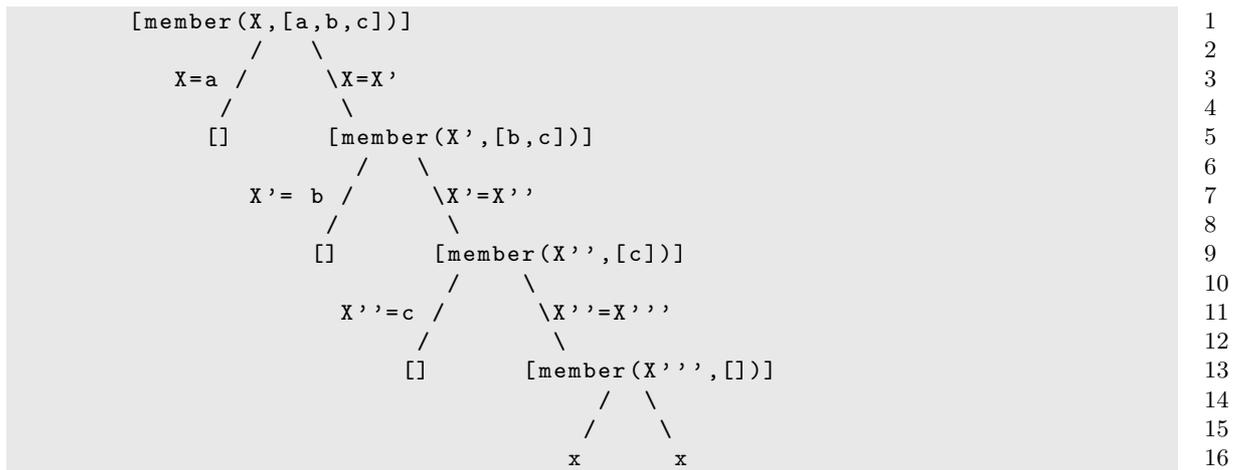
\begin{lstlisting}[style=search]
        [member(X,[a,b,c])]
                /   \
           X=a /     \X=X'
              /       \ 
             []      [member(X',[b,c])]
                       /   \
                X'= b /     \X'=X''
                     /       \
                    []      [member(X'',[c])]
                             /    \
                      X''=c /      \X''=X'''
                           /        \
                          []      [member(X''',[])]
                                       /  \  
                                      /    \ 
                                     x      x
\end{lstlisting}

The result of this query are three solutions: \verb+X=a+, \verb+X=b+ (because \verb+X=X'+ and \verb+X'=b+), and \verb+X=c+ (because \verb+X=X'+, \verb+X'=X''+ and \verb+X''=c+).

\section{Search Trees with Negation and the Cut}\label{sec:negation}

\subsection{The built-in control predicates}

There are a couple of built-in predicates that affect the way search trees
are constructed. One important one is the cut. In order to get good understanding of the cut,
we first need to look at the logical predicates.
The predicate \texttt{true/0} always succeeds; the predicates
\texttt{fail/0} (or the equivalent \texttt{false/0}) always fail. Hence, their
corresponding search trees are as follows:

\begin{figure}[h]
\begin{lstlisting}[style=rule]
  [fail]     [fail,Q]     [false]     [false,Q]    [true]     [true,Q]
    |           |            |            |           |           |
    |           |            |            |           |           |
    x           x            x            x          []          [Q]
\end{lstlisting}
\caption{Search tree expansion rules for the logical predicates.}
\end{figure}

\subsection{The cut}

Without doubts, one of the most interesting (and one of the most
trickiest) tools of Prolog is the \textit{cut} predicate.  It is a
predicate that always succeeds. But it has an extremely useful
side-effect: it cuts open branches from the search tree, thereby
giving direct control on memory management. Cuts are used to make
Prolog programs efficient. This often goes at the cost of making
Prolog programs less declarative and readable. The fact or the matter
is that cuts are needed in practical applications of logic
programming.

The cut is not an easy concept to understand. Some newcomers to Prolog
confuse \textit{cut} with a \textit{halt} or a \textit{stop}. But it
has little to do with that, because it always succeeds. What \textit{cut} does is pruning
open branches (choice points) in the search tree. But, and here is where it gets tricky, not necessarily \textit{all} open
branches in the search tree. The effect of the cut predicate is to
discard all choice points (open branches) created since entering the
predicate in which the cut appears.  In other words, commit to the
clause (rule) in which the cut appears and discard choice points that
have been created by goals to the left of the cut in the current
clause. So the general schemata look like:

\begin{figure}[h]
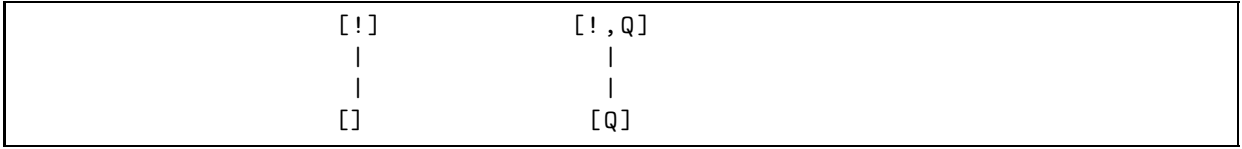

\begin{lstlisting}[style=rule]
                   [!]           [!,Q]
                    |              |
                    |              |
                   []             [Q]
\end{lstlisting}
\caption{Search tree expansion rules for the cut.}
\end{figure}

But these rules don't show anything about cutting branches. So let's have a look at an example using
 cut.

\subsection{An example using cut}

Consider the following Prolog database:

\begin{lstlisting}[style=prolog,caption=Prolog database to illustrate the cut]
a(X):- b(X).          
a(X):- c(X).          

b(X):- d(X), !.       

b(X):- e(X).          

c(1).
c(2).

d(3).
d(4). 

e(5).
\end{lstlisting}

This Prolog database defines four predicates: \texttt{a/1},
\texttt{b/1}, \texttt{c/1}, and \texttt{d/1}, using three rules and
four facts. It doesn't have any practical usage, except from
explaining how the cut works. Now consider the query \texttt{a(A)}.
As there are two rules for \texttt{a/1}, we can draw two branches:

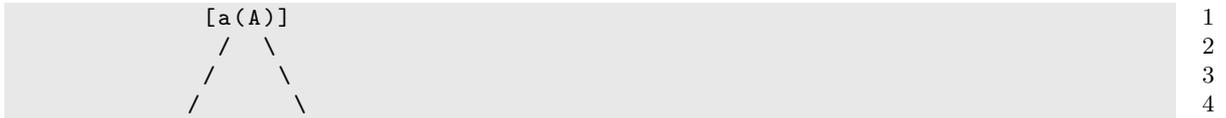
\begin{lstlisting}[style=search]
             [a(A)]
              /  \
             /    \
            /      \
\end{lstlisting}

The first branch corresponds to a rule, introducing a new query:

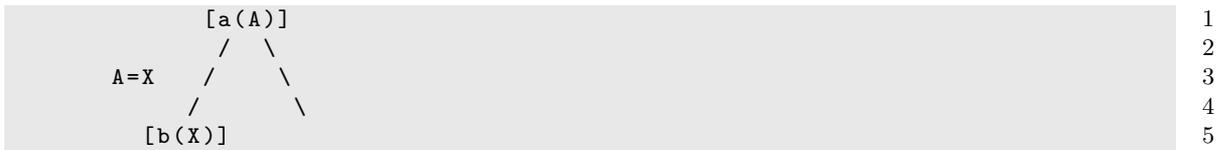
\begin{lstlisting}[style=search]
             [a(A)]
              /  \
       A=X   /    \
            /      \
         [b(X)]
\end{lstlisting}

This query can be satisfied by two clauses, so we create two new branches, and concentrate on the first one. This is a query with a cut:

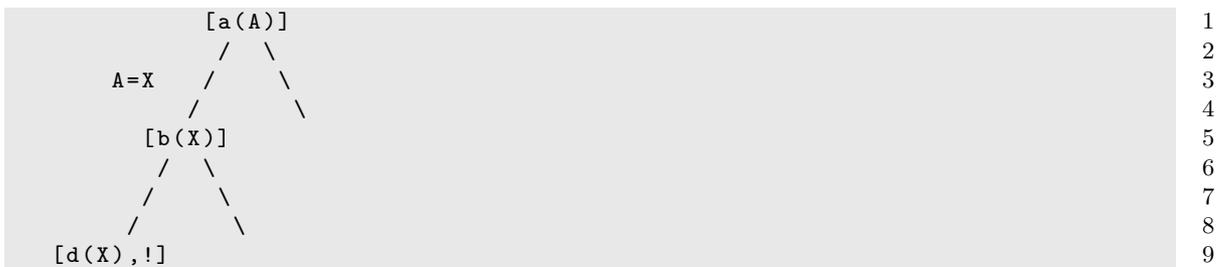
\begin{lstlisting}[style=search]
             [a(A)]
              /  \
       A=X   /    \
            /      \
         [b(X)]
          /  \
         /    \
        /      \
   [d(X),!]
\end{lstlisting}

As this is a compound, conjunctive query (the second part just happens
to be a cut), we first look at the left conjunct, the query \texttt{d(X)}. There are two possibilities to satisfy this query, so we draw
two branches. The first branch succeeds, and the variable \texttt{X}
is unified with atom \texttt{3}. And now we have to deal with the cut:

\begin{lstlisting}[style=search]
             [a(A)]
              /  \
       A=X   /    \
            /      \
         [b(X)]
          /  \
         /    \
        /      \
   [d(X),!]
     /  \
X=3 /    \
   /      \
 [!]
\end{lstlisting}

This cut, as all cuts do, succeeds. But it cuts down all open branches
that were created since it was introduced. There are three open
branches in the tree above. The first open branch was introduced
before the cut came in to play, so it is not affected (just check the
Prolog database to verify this). What about the second branch? This
branch was triggered by the predicate \texttt{b/1} that also introduced the cut.
As a consequence, this branch is pruned away from the search tree, as is the third open branch.
This gives the following tree, with the two branches removed (signalled by \texttt{X}):

\begin{lstlisting}[style=search]
             [a(A)]
              /  \
       A=X   /    \
            /      \
         [b(X)]
          /  \
         /    X
        /      \
   [d(X),!]
     /  \
X=3 /    X
   /      \
 [!]
  |
  |
 []
\end{lstlisting}

At this point, backtracking will jump to the one open branch that is
left, corresponding to the second possibility for satisfying the query \texttt{a(A)}.
The final search tree, then, looks like this:

\begin{lstlisting}[style=search]
             [a(A)]
              /  \
       A=X   /    \ A=X
            /      \
         [b(X)]    [c(X)]
          /  \       |  \
         /    X   X=1|   \ X=2
        /      \     |    \
   [d(X),!]         []    []
     /  \
X=3 /    X
   /      \
 [!]
  |
  |
 []
\end{lstlisting}

In sum, there are three solutions for this query. Without the cut in
the program there would have been five answers.

\subsection{Negation as failure}

In Prolog, a negated query \texttt{Q} is written as \texttt{not(Q)} or
\verb=\+ Q=.  A negated guery succeeds if an attempt to falsify it
fails \citep{clocksinmellish:prolog}.  This is called \textit{negation
  as failure} \citep{bratko}.  Prolog's negation makes use of the cut
and the built-in predicate \texttt{fail/0}.  When used, it creates two
new branches in the tree. In the first branch the query in the scope
is placed followed by the cut and \texttt{fail/0}. The second branch
succeeds (but it will be cut from the tree in case the query of the
first branch succeeds). This is how the search tree expansion rule for
negation looks like:

\begin{figure}[h]
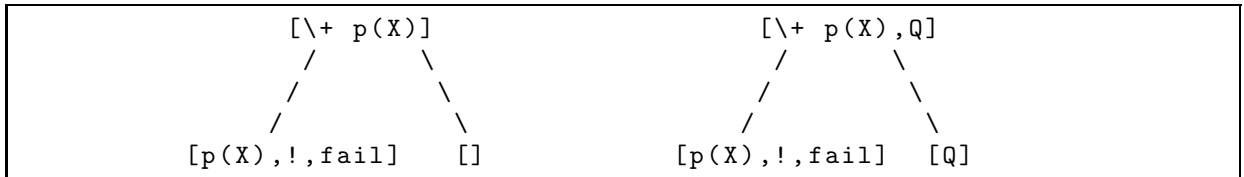

\centering
\begin{lstlisting}[style=rule]
                [\+ p(X)]                   [\+ p(X),Q]
                 /      \                    /      \
                /        \                  /        \
               /          \                /          \
          [p(X),!,fail]   []           [p(X),!,fail]  [Q]
\end{lstlisting} 
\caption{Search tree expansion rules for negation.}
\label{fig:rulenegation}
\end{figure}

The expansion rules in Figure~\ref{fig:rulenegation} also show why
variables in the scope of a Prolog negation are never bound. This is
because Prolog negation introduces two branches, and variables on
different branches are independent of each other. So even if the query \textbf{Q} in
in Figure~\ref{fig:rulenegation} would contain an occurrence of the variable \textbf{X}, it would be effected by any bindings that were made while trying to proof the query \texttt{p(X)}.

\bigskip

Now let's look at an example. We have three children (Ann, Ben and
Cai) with their ages stored in a Prolog database. And we have defined
a predicate that computes who is the youngest child 
(Listing~\ref{listing:youngest}).

\bigskip

\begin{lstlisting}[style=prolog,label=listing:youngest,caption=A Prolog database and definition for youngest/1.]
age(ann,5).           
age(ben,4).
age(cai,6).         

youngest(X):- age(X,Y), \+ (age(_,Z), Z < Y).
\end{lstlisting}

Here, in line 5 of Listing~\ref{listing:youngest},
``the youngest'' is defined as a child with a certain age \texttt{Y},
such that there is no child that has an age that is smaller than
\texttt{Y}. 
 Well, who is the youngest? We can find out by drawing the search tree for \texttt{youngest(Who)}. 
There is only one rule for \texttt{youngest/1}, so we draw just one branch with a complex query that is
formed by the body of the rule:

\begin{lstlisting}[style=search]
     [youngest(Who)]
          |
          | Who=X
          |
    [age(X,Y), \+ (age(_,Z), Z < Y)]
\end{lstlisting}

This is a compound query, with the first conjunct a simply query with
variables, and the second conjunct a negated query of another compound
query.  We look at the left conjunct first. There are three facts for
the predicate \texttt{age/2} in the database (line 1--3 in
Listing~\ref{listing:youngest}), so we draw three branches. We look at
the first branch first: perhaps Ann with the age of 5 is the youngest.

\clearpage

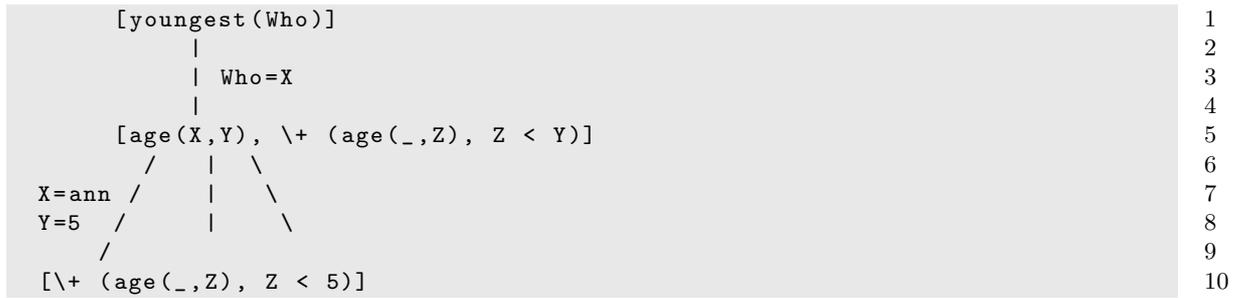
\begin{lstlisting}[style=search]
       [youngest(Who)]
            |
            | Who=X
            |
       [age(X,Y), \+ (age(_,Z), Z < Y)]
         /   |  \
  X=ann /    |   \
  Y=5  /     |    \
      /
  [\+ (age(_,Z), Z < 5)]
\end{lstlisting}

Now we deal with the negation, using the search tree expansion rule
for negation, introducing the cut-fail combination:

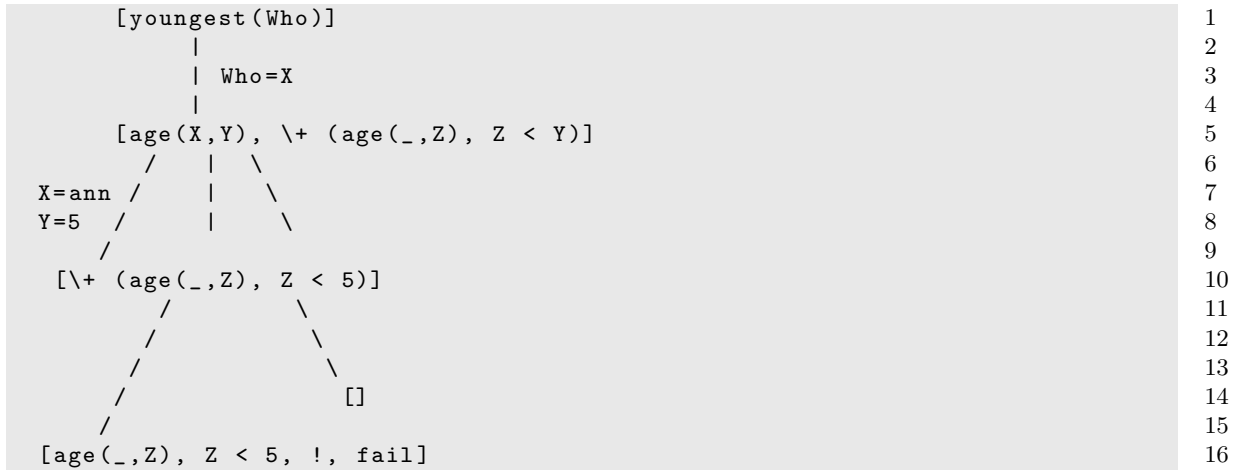
\begin{lstlisting}[style=search]
       [youngest(Who)]
            |
            | Who=X
            |
       [age(X,Y), \+ (age(_,Z), Z < Y)]
         /   |  \
  X=ann /    |   \
  Y=5  /     |    \
      /      
   [\+ (age(_,Z), Z < 5)]
          /        \
         /          \
        /            \
       /              []
      /
  [age(_,Z), Z < 5, !, fail]    
\end{lstlisting}

Once again, for \texttt{age/3} we have three possibilities, so we draw three branches. The first branch fails (because Ann is not younger than Ann), the second intially succeeds (Ben is younger than Ann), but then the cut removes the two open branches:

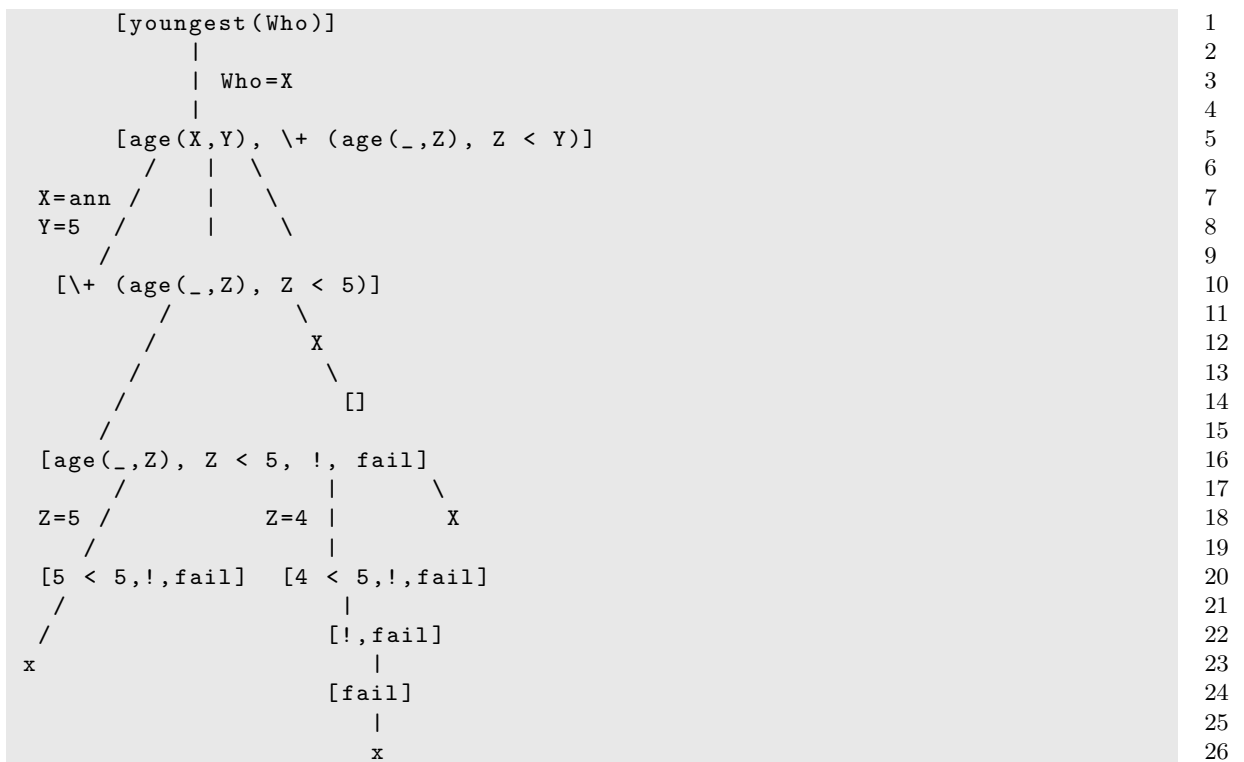
\begin{lstlisting}[style=search]
       [youngest(Who)]
            |
            | Who=X
            |
       [age(X,Y), \+ (age(_,Z), Z < Y)]
         /   |  \
  X=ann /    |   \
  Y=5  /     |    \
      /      
   [\+ (age(_,Z), Z < 5)]
          /        \
         /          X
        /            \
       /              []
      /
  [age(_,Z), Z < 5, !, fail]    
       /             |      \
  Z=5 /          Z=4 |       X
     /               |        
  [5 < 5,!,fail]  [4 < 5,!,fail]
   /                  |
  /                  [!,fail]
 x                      |
                     [fail]
                        | 
                        x 
\end{lstlisting}

No, Ann is not the youngest. The cut was activated and removed two of the open branches. The \texttt{fail/0} predicates ensures that we look at the two other possibilities.
Perhaps Ben or Cai is (we don't redraw the part 
of the search tree for reasons of limited space):

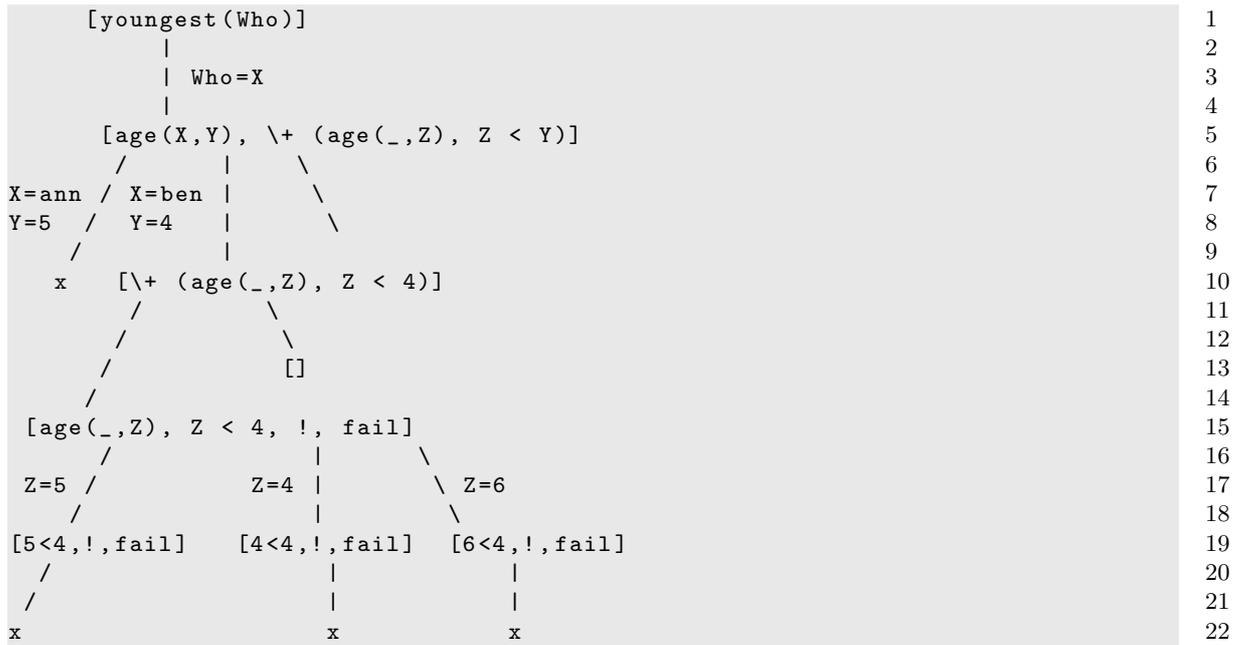
\begin{lstlisting}[style=search]
     [youngest(Who)]
          |
          | Who=X
          |
      [age(X,Y), \+ (age(_,Z), Z < Y)]
       /      |    \
X=ann / X=ben |     \
Y=5  /  Y=4   |      \
    /         |      
   x   [\+ (age(_,Z), Z < 4)]
        /        \
       /          \
      /           []
     /
 [age(_,Z), Z < 4, !, fail]    
      /             |      \
 Z=5 /          Z=4 |       \ Z=6
    /               |        \ 
[5<4,!,fail]   [4<4,!,fail]  [6<4,!,fail]
  /                  |           |
 /                   |           |
x                    x           x
\end{lstlisting}

Yes, we have a positive answer. The cut is not activated, and the top
query succeeds with variable \texttt{Who} unified with atom
\texttt{ben}. There is still one open branch, but backtracking will
not lead to more solutions here.

\clearpage
\section{Search Trees in Learn Prolog Now!} \label{sec:lpn}

Some of the search trees that you find in \textit{Learn Prolog Now!} 
\citep{lpn,ptds} are incomplete, or are not drawn in the way shown in this manual.
We will present them here in the order they appear in the book.

\subsection{Proof Search, Chapter 2 from Learn Prolog Now!}

In this chapter of Learn Prolog Now, proof search is illustrated with the following knowledge base \cite[p.34]{lpn}:

\bigskip

\begin{lstlisting}[style=prolog,label=listing:proof,caption=Prolog database to illustrate proof search.]
f(a).  
f(b).  

g(a).  
g(b).

h(b).

k(X):- f(X), g(X), h(X).
\end{lstlisting}

The query posed is \texttt{k(Y)}. There is only one answer. The way
Prolog works this out is illustrated by the following search tree:

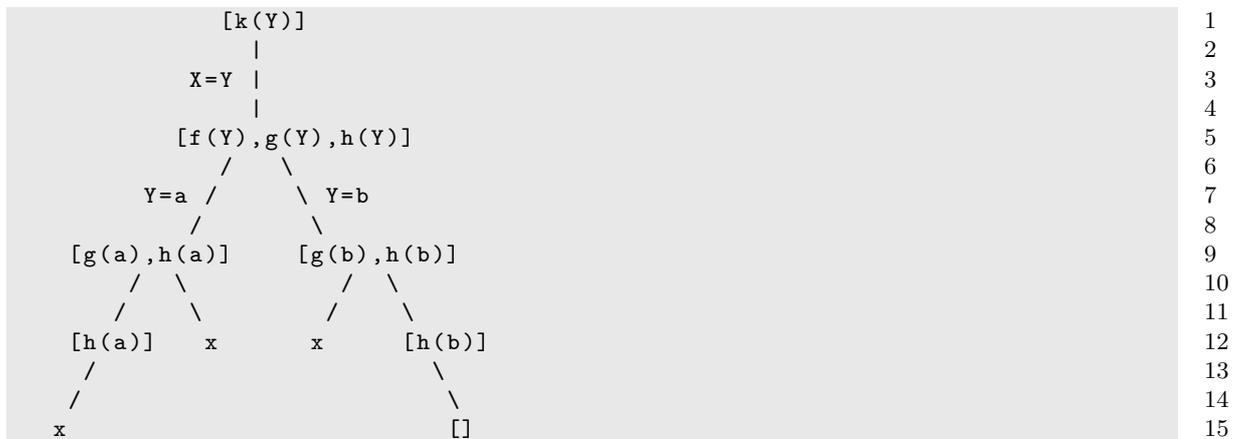
\begin{lstlisting}[style=search]
              [k(Y)]
                |
            X=Y |
                |
           [f(Y),g(Y),h(Y)]
              /   \
         Y=a /     \ Y=b
            /       \
    [g(a),h(a)]    [g(b),h(b)]
        /  \          /  \
       /    \        /    \
    [h(a)]   x      x     [h(b)]
     /                      \
    /                        \
   x                         []
\end{lstlisting}

Even though in \textit{Learn Prolog Now!} the tree is shown with
variable renaming, this is not required for this example. Another
difference with \textit{Learn Prolog Now!} are the two branches that we draw for
the query \texttt{g(a)}. Here it is correct to draw two branches
following our rules for drawing search trees, as the predicate \texttt{g/1} is
defined by two clauses. Having said that, it is also clear that the
second branch fails immediately, and in cases like these we could
consider some flexibility and be less strict in showing all search
possibilities in the tree, leaving out obvious pre-mature dead branches.

\clearpage

\subsection{Recursion, Chapter 3 from Learn Prolog Now!}

The search tree for the query \texttt{descent(anne,donna)} in
Learn Prolog Now! is incomplete \cite[p.54]{lpn}. The search tree requires
variable renaming for the variable \texttt{Z}, and we do so by
introducing \texttt{Z1}, \texttt{Z2}, and so on. The complete search
tree is enormous, but we managed to get it on one page:

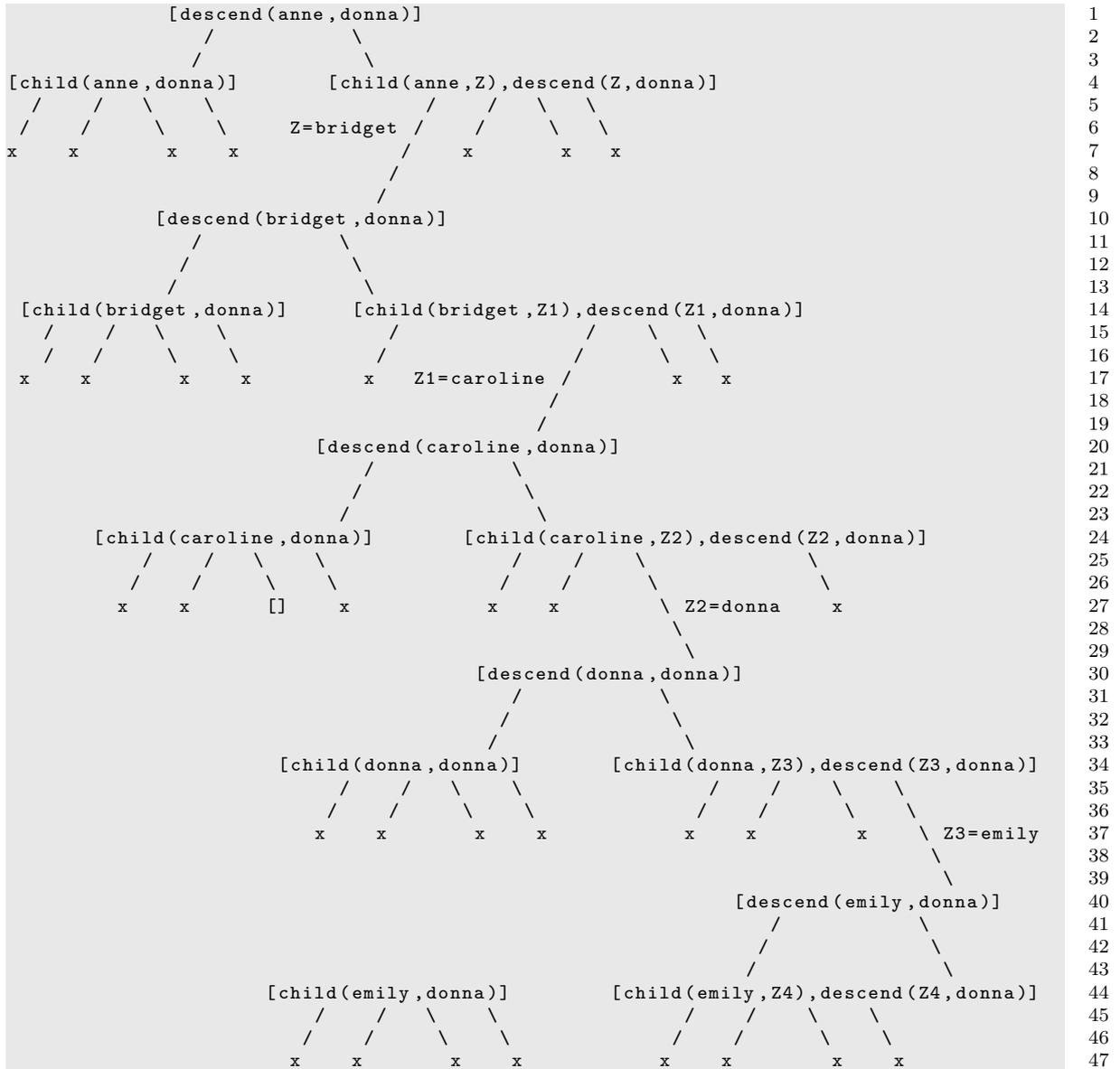
\begin{lstlisting}[style=search,basicstyle=\ttfamily\scriptsize]
             [descend(anne,donna)]
                /           \
               /             \
[child(anne,donna)]       [child(anne,Z),descend(Z,donna)]
  /    /   \    \                 /    /   \   \
 /    /     \    \     Z=bridget /    /     \   \
x    x       x    x             /    x       x   x
                               /
                              /
            [descend(bridget,donna)]
               /           \
              /             \ 
             /               \
 [child(bridget,donna)]     [child(bridget,Z1),descend(Z1,donna)]
   /    /   \    \             /               /    \   \
   /   /     \    \           /               /      \   \ 
 x    x       x    x         x   Z1=caroline /        x   x
                                            /
                                           /
                         [descend(caroline,donna)]
                             /           \
                            /             \ 
                           /               \
       [child(caroline,donna)]       [child(caroline,Z2),descend(Z2,donna)]
           /    /   \    \               /    /    \             \
          /    /     \    \             /    /      \             \
         x    x      []    x           x    x        \ Z2=donna    x
                                                      \
                                                       \
                                      [descend(donna,donna)]
                                         /           \
                                        /             \ 
                                       /               \
                      [child(donna,donna)]       [child(donna,Z3),descend(Z3,donna)]
                           /    /   \    \               /    /    \    \               
                          /    /     \    \             /    /      \    \
                         x    x       x    x           x    x        x    \ Z3=emily  
                                                                           \ 
                                                                            \
                                                           [descend(emily,donna)]
                                                              /           \
                                                             /             \ 
                                                            /               \
                     [child(emily,donna)]        [child(emily,Z4),descend(Z4,donna)]
                         /    /   \    \               /    /    \    \               
                        /    /     \    \             /    /      \    \
                       x    x       x    x           x    x        x    x
\end{lstlisting}

Note that there is only one successful branch, shown in line 27 with the empty box.

\subsection{Recursion, Chapter 3 from Learn Prolog Now!}

Example~4 of Chapter~3 in Learn Prolog Now! is about the recursive
predicate \texttt{add/3} \cite[p.57]{lpn}, which we give here again in
Listing~\ref{listing:add}.

\begin{lstlisting}[style=prolog,label=listing:add,caption=Prolog definition for add/3.]
add(0,Y,Y).
add(succ((X),Y,succ(Z)):- add(X,Y,Z).
\end{lstlisting}

The search tree given two pages later is incomplete \cite[p.59]{lpn}. Here is the complete search tree:

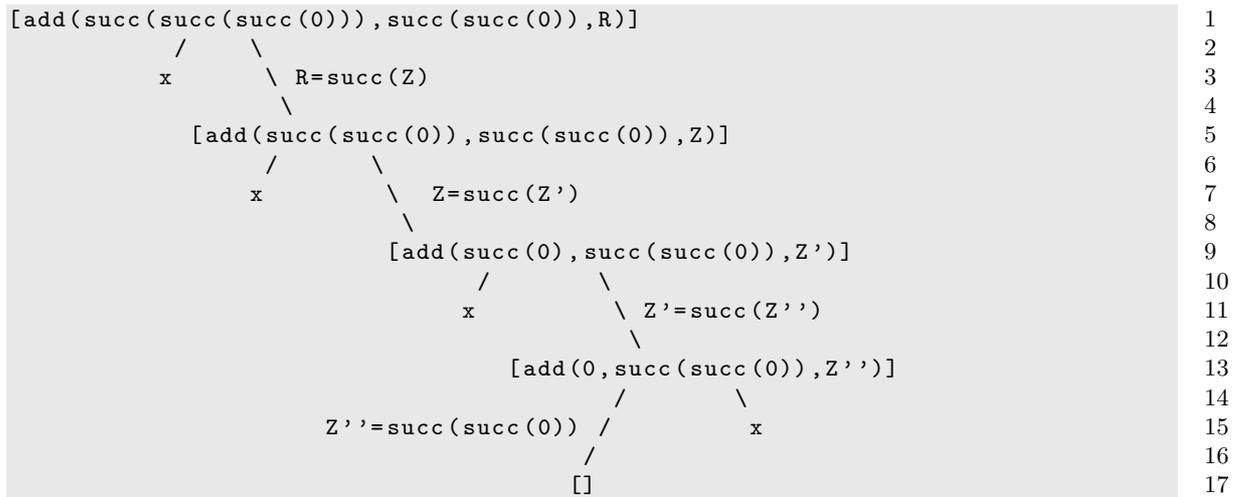
\begin{lstlisting}[style=search]
[add(succ(succ(succ(0))),succ(succ(0)),R)]
           /    \
          x      \ R=succ(Z)
                  \
            [add(succ(succ(0)),succ(succ(0)),Z)]
                 /      \
                x        \  Z=succ(Z')
                          \
                         [add(succ(0),succ(succ(0)),Z')]
                               /       \
                              x         \ Z'=succ(Z'')
                                         \
                                 [add(0,succ(succ(0)),Z'')]
                                        /       \
                     Z''=succ(succ(0)) /         x
                                      /
                                     []  
\end{lstlisting}

The query succeeds, and the variable \texttt{R} unifies with \texttt{succ(succ(succ(succ(succ(0)))))}.

\vfill

\subsection{More Lists, the predicate append/3}

In Chapter 6 of \textit{Learn Prolog Now!}, the idea behind the
standard Prolog predicate \texttt{append/3} is explained \cite[p.106]{lpn}. The
recursive definition provided contains two clauses (a fact and a rule):

\begin{lstlisting}[style=prolog,caption=Definition of append/3.]
append([],L,L).
append([H|T],L2,[H|L3]):- append(T,L2,L3).
\end{lstlisting}

 This is done with the help of a search
tree 
for the query \texttt{append([a,b,c],[1,2,3],X)}
a couple of pages later.
The search tree is incomplete: it only shows the
succesful branches \cite[p.108]{lpn}.
The complete search tree is realised as follows:

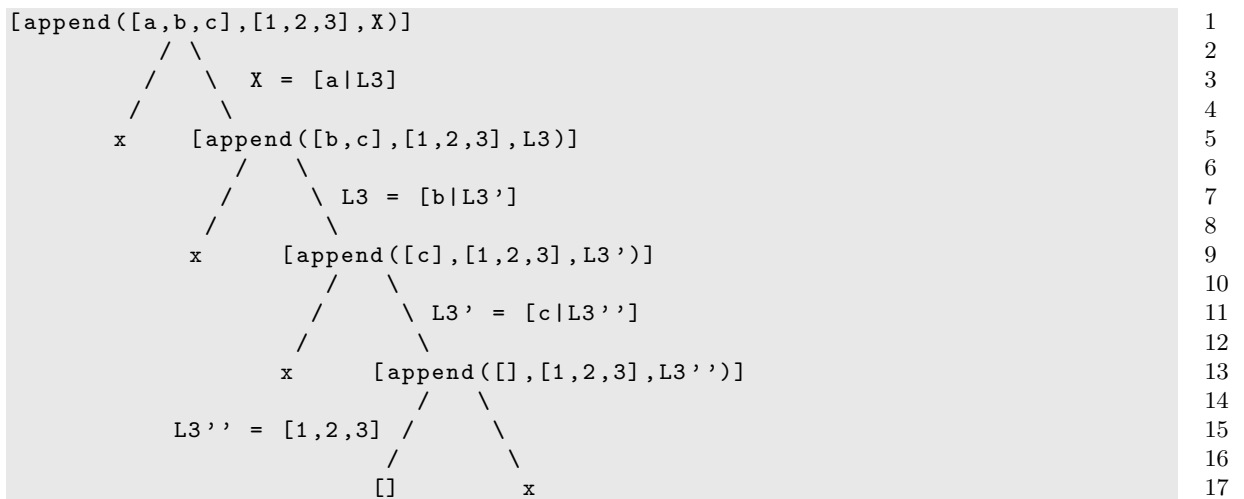
\begin{lstlisting}[style=search]
[append([a,b,c],[1,2,3],X)]
          / \
         /   \  X = [a|L3]
        /     \     
       x    [append([b,c],[1,2,3],L3)]
               /   \
              /     \ L3 = [b|L3']
             /       \
            x     [append([c],[1,2,3],L3')]
                     /   \
                    /     \ L3' = [c|L3'']
                   /       \
                  x     [append([],[1,2,3],L3'')]
                           /   \
           L3'' = [1,2,3] /     \
                         /       \
                        []        x
\end{lstlisting}

There is one closed branch, so Prolog will answer in the affirmative
way and unify the variable \texttt{X} with the list
\texttt{[a,b,c,1,2,3]}.

\clearpage
\subsection{Search tree for the last query of Exercise 2.2 from Learn Prolog Now!}

In Exercise 2.2 we are asked to draw the search tree for \texttt{magic(Hermione)}. The answer given is correct \cite[p.239]{lpn} but it consists of unnecessary variable renaming. A simplified search tree is:

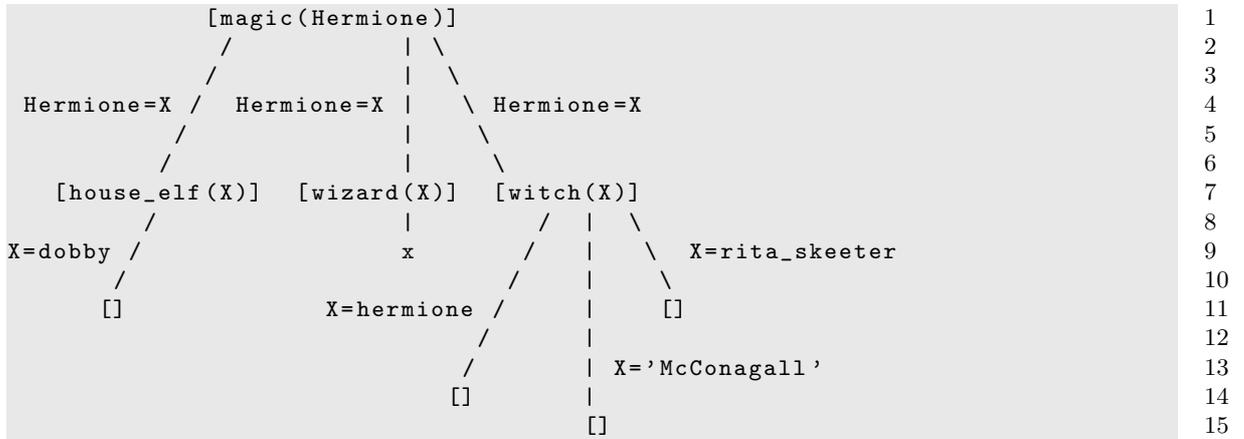
\begin{lstlisting}[style=search]
             [magic(Hermione)]
              /           | \
             /            |  \ 
 Hermione=X /  Hermione=X |   \ Hermione=X
           /              |    \
          /               |     \
   [house_elf(X)]  [wizard(X)]  [witch(X)]
         /                |        /  |  \
X=dobby /                 x       /   |   \  X=rita_skeeter
       /                         /    |    \
      []             X=hermione /     |    []
                               /      |     
                              /       | X='McConagall'
                             []       |
                                      []
\end{lstlisting}

\subsection{Complete answers to Exercise 4.7 from Learn Prolog Now!}

The answers to Exercise 4.7, drawing three search trees for the predicate \texttt{member/3}, are all incomplete \cite[p.244]{lpn}.
Here are the complete search trees:

\begin{lstlisting}[style=search]
        [member(a,[c,b,a,y])]
                /   \
               /     \
              /       \ 
             x     [member(a,[b,a,y])]
                       /   \
                      /     \
                     /       \
                    x      [member(a,[a,y])]
                             /    \
                            /      \
                           /        \
                          []      [member(a,[y])]
                                     /  \  
                                    /    \ 
                                   x     [member(a,[])]
                                             /  \
                                            /    \
                                           x      x
\end{lstlisting}

\begin{lstlisting}[style=search]
        [member(x,[a,b,c])]
                /   \
               /     \
              /       \ 
             x     [member(x,[b,c])]
                       /   \
                      /     \
                     /       \
                    x      [member(x,[c])]
                             /    \
                            /      \
                           /        \
                          x      [member(x,[])]
                                       /  \  
                                      /    \ 
                                     x      x
\end{lstlisting}

\begin{lstlisting}[style=search]
        [member(X,[a,b,c])]
                /   \
           X=a /     \X=X'
              /       \ 
             []      [member(X',[b,c])]
                       /   \
                X'= b /     \X'=X''
                     /       \
                    []      [member(X'',[c])]
                             /    \
                      X''=c /      \X''=X'''
                           /        \
                          []      [member(X''',[])]
                                       /  \  
                                      /    \ 
                                     x      x
\end{lstlisting}

\vfill

\bibliographystyle{chicago}
\bibliography{prolog}

\end{document}